# Comment about the use of Monte-Carlo methodology for the representation of atomic electronic densities

Jean Hoarau and Jean-Claude Rayez*
Université de Bordeaux – Institut des Sciences Moléculaires (ISM - CNRS) UMR 5255
33405 Talence Cedex – France

___________________________________________________________________

ABSTRACT : Representations of atomic orbitals based on Monte-Carlo (MC) approaches are not always correct when using various sets of orthogonal coordinates other than Cartesian coordinates. The analysis proposed here gives elements for a proper use of MC methodology. It can be very useful for students and for teachers.

___________________________________________________________________

## Introduction

Graphic representations of atomic orbitals are now largely used in courses of quantum chemistry since the pioneering work of D.T. Cromer (1). The fantastic development of computing tools allows very suggestive descriptions of atomic and molecular objects for educational purposes. Monte-Carlo (MC) approach is a very elegant one to reach such objectives. Many articles have been published on this subject since more than forty years, but few of them deal specifically with MC methodology.

The object of this article is to emphasize a difficulty which appears when using MC method for any set of orthogonal coordinates other than Cartesian coordinates. Mathematically, such a situation corresponds to orthogonal coordinates for which the elementary volume dv exhibits a scalar (due to the orthogonality property) Jacobian factor J different of unity. For instance, for spherical coordinates $dv = J\, dr\, d\theta\, d\varphi$ with $J = r^2\sin\theta$. In the case of Cartesian coordinates, $J = 1$. Such a situation leads to some caution for the use of MC.

This problem has already been largely treated in Mathematics. It exists a large bibliography about it. Let us cite some of these documents which can be obtained now directly from the Web (2). Nevertheless, all the physical chemists are not fully familiar with such a topic. Therefore, it turns out that this problem of using non Cartesians coordinates is either ignored in some textbooks or articles or, if it is known, the authors do not consider it sufficiently important to be pointed out.

## The problem

Let us first refresh our mind about Monte-Carlo method. The hydrogen nucleus being at the origin of the frame, we select randomly a point in the plane (x, y). Here $z = 0$. The extension to a 3D representation, as done by Cromer, is rather straightforward (see next section). The values of the square $\psi^2_{n,l,m}$ of a hydrogenlike wave function $\psi_{n,l,m}$ and the maximum value of $\psi^2_{n,l,m}$ : $\psi^2_{max}$ are calculated at the point (x, y). The ratio $\omega = \psi^2/\psi^2_{max}$ is then compared to another random number $\xi$ belonging to the interval [0, 1]. If $\omega > \xi$, a dot is plotted in the plane (x, y). Otherwise it is not. As written by Cromer (1), this procedure, which makes the density of points in the (x, y) plane proportional to $\psi^2$, is repeated until the desired number of points is plotted. In this 2D case, $\psi^2$ is a probability by surface unit.

Using Cartesian coordinates x, y, we select the same number of x and y values. Of course, it does not come to the mind to use different numbers. In other words, we want a uniform distribution of points in the (x, y) plane. With such a 2D random space, the electronic density associated with any atomic function will be correctly reproduced.

When using polar coordinates r and θ, the Jacobian factor of the elementary 2D "volume" rdrdθ is equal to r instead of 1. Then, selecting the same number of r and θ values does not lead to a uniform (random) distribution of points in the plane. As a matter of fact, if we consider a set of narrow annular strips of increasing radius belonging to a circular disk, an increasing number of points in each annular strip is necessary to get a uniform density of dots as long as the radius grows.

The problem is how do we determine the number of θ values for a given number of r values? The problem is not totally trivial. As a matter of fact, discussing with some teachers, they did not raised up this problem, taking *a priori* the same number of r and θ values. Then; we suspect that some of the (r, θ) Monte-Carlo drawings already published are not correct.

## Solution and discussion

As we wrote in the introduction, it exists a significant bibliography to solve this problem. Here, we give another presentation of its solution.

Let us define $N(r_{max})$ the total number of r values selected randomly between 0 and a maximum value $r_{max}$ determined by a given threshold of electronic density. $N_r(\theta)$ is the number of θ values chosen randomly between 0 and 2π for a given r value belonging to the domain [0, $r_{max}$]. A simple way to simulate a random distribution of n points on a segment of length L is to take n equivalent intervals on L and put one point in each interval. The distance between two consecutive points is L/n. Then, the average gaps Δr and Δθ between two closest values of r and θ are:

$$\Delta r = \frac{r_{max}}{N(r_{max})} \qquad \text{and} \qquad \Delta\theta = \frac{2\pi}{N_r(\theta)} \qquad [1]$$

To get a uniform 2D distribution of points, we have to associate with each point a small area Δr . rθ which must be a "squared" area (the word "squared" is placed between quotation marks since rΔθ is not really a straight segment but rather a circular segment). Then, Δr should be equal to rΔθ. Another argument can be based on the expression of the differential vector d**r** in polar coordinates. Since **r** = r.**u**, where **u** is the unitary vector (sometimes noted $\hat{\mathbf{r}}$ in some text books) along **r**, d**r** = dr.**u** + r.dθ.**v** with **v** the unitary vector (also noted $\hat{\theta}$ ) along a direction orthogonal to **u**. The displacements along **u** and **v** are respectively dr and rdθ.

Imposing Δr = rΔθ, it comes out from [1]: $N_r(\theta) = N(r_{max}).\dfrac{2\pi.r}{r_{max}}$ [2]

To check the validity of this analysis, we have performed three kinds of calculations:

i)      We start with a disk of area 100Å$^2$. The radius of this disk is: $r_{max} = (100/\pi)^{1/2} = 5.642$Å. We select $N(r_{max}) = 100$ random r points between 0 and $r_{max}$. According to relation [2]: $N_{r_{max}}(\theta) = 2\pi.100 = 628$ since we must keep an integer number. With all the $N_r(\theta)$ integer values, we obtained numerically the total number of points selected in the disk : $\sum_{r=0}^{r_{max}} N_r(\theta) = 31714$. In this expression, do not forget that $N_r(\theta)$ is just an integer number, not a function of $\theta$. See the footnote (3) in which we calculate the magnitude of this summation analytically. Figure 1a displays the uniform distribution in the disk.

ii)      Using the same disk, we take now the same number of points for r and for $\theta$: 100 r points and 100 $\theta$ points (instead of sets of $\theta$ values from 6 (the integer part of 6.28) to 628 values that is to say a total of 31714 values). Then we deal now with a selection of 10000 points instead of 31714. This difference does not play any role for our analysis, but this choice leads to a clearer picture than with a larger set of points. Figure 1b shows the non uniform distribution of points, with a large concentration near the centre of the disk and a decreasing one as long as we move to the periphery of the disk.

iii)      The analysis in Cartesian coordinates is performed in a square of side $2.r_{max}$. It is the square in which the disk is fully included. Since the square has a larger area than the disk, we have taken $31714 \times 4/\pi = 40380$ points from the ratio of both surfaces. Figure 1c displays the uniform distribution of points obtained from a random selection of x and y coordinates.

Using these three distributions to represent the electronic density of the normalized hydrogen function $2p_x$ ($a_0 = 0.529$ Å) :

$$\psi_{2p_x} = \frac{1}{4\sqrt{2\pi}} \left(\frac{1}{a_0}\right)^{5/2} \exp\left[-r/2a_0\right] r \sin\theta\cos\varphi$$

$$= \frac{1}{4\sqrt{2\pi}} \left(\frac{1}{a_0}\right)^{5/2} \exp\left[-(x^2+y^2+z^2)^{1/2}/2a_0\right] x ,$$

we obtained the three following representations :

i) Figure 2a based on a uniform scanning of the circular disk (from Figure 1a).
ii) Figure 2b using the non uniform exploration of the disk (from Figure 1b),
iii) Figure 2c using random Cartesian coordinates (from Figure 1c).

It is clear that Figures 2a and 2c are very similar, the small difference being due to the use of a larger set of points (40380) in the square than in the disk (31714). On the opposite, Figure 2b is strongly different from the two others with a much darker central region than near the periphery, in accord with the non uniform scanning of the disk displayed in Figure 1b.
*A priori*, if the comparison between Figure 1a and 1b is not in front your eyes or kept in mind, it is not obvious that the representation displayed in Figure 2b is wrong and the one shown in Figure 2a or 2c is correct. For such a function, it is their symmetry properties, like symmetry and nodal planes, which appear immediately at the first glance, not the fact that the density near the H nucleus is over represented.
Now, you have two MC representations of 2p functions: Figure 2a or 2c and Figure 2b. In all the text books you can have in yours hands, look at the MC representations (if there are some) of any 2p function and make your mind. You should be now a good critic of what you see.

**Extension to a spherical representation and to other sets of 3D orthogonal coordinates**

i) For 3D spherical coordinates, the elementary volume is $dv = r^2 \sin\theta.drd\theta d\varphi$ and the differential vector $d\mathbf{r} = dr\ \mathbf{u} + r.d\theta\ \mathbf{v} + r.\sin\theta.d\varphi\ \mathbf{w}$. where $\mathbf{u}$, $\mathbf{v}$ and $\mathbf{w}$ are the three unitary vectors (orthogonal each other) associated with the spherical coordinates.

The following relations between the average variations $\Delta r$, $\Delta\theta$, $\Delta\varphi$ are the following :

$$\Delta r = \frac{r_{max}}{N(r_{max})} \qquad \Delta\theta = \frac{\pi}{N_r(\theta)} \qquad \Delta\varphi = \frac{2\pi}{N_{r,\theta}(\varphi)}$$

where $N(r_{max})$ is the number of r points selected randomly between 0 and the radius $r_{max}$ which becomes now the radius of a sphere, $N_r(\theta)$ is the number of $\theta$ values selected randomly between 0 and $\pi$ for a given value of r and $N_{r,\theta}(\varphi)$ is the number of $\varphi$ values selected randomly between 0 and $2\pi$ for given values of r and $\theta$.
As for the 2D analysis, a uniform distribution in a spherical space will be obtained if the elementary volume is a "cube". Then :

$$\Delta r = r . \Delta\theta = r.\sin\theta . \Delta\varphi$$

From these equalities, we get :

$$N_{r,\theta}(\varphi) = N_r(\theta).2\sin\theta = N(r_{max}).\frac{\pi.r}{r_{max}}$$

The procedure is immediate. We selected randomly $N(r_{max})$ points between 0 and $r_{max}$. For each r value, we choose randomly $N_r(\theta)$ points in $\theta$ between 0 and $\pi$ according to the second equality above. Then for given r and $\theta$ values, we selected randomly $N_{r,\theta}(\varphi)$ points in $\varphi$ between 0 and $2\pi$ according to the first equality. The result of this work shows a sphere uniformly full of dots.
*Remark :* There is no matter to label some specific coordinates in the order 1, 2, 3. Using spherical coordinates, we chose r as coordinate 1, $\theta$ as coordinate 2 and $\varphi$ as coordinate 3. Of course we could have selected any other choice.

ii) For any 3D set of orthogonal coordinates $q_1$, $q_2$, $q_3$, the volume element is $dv = h_1 h_2 h_3 q_1 q_2 q_3$, the $h_i$'s being defined as (4) :

$$h_i^2 = \left(\frac{\partial x}{\partial q_i}\right)^2 + \left(\frac{\partial y}{\partial q_i}\right)^2 + \left(\frac{\partial z}{\partial q_i}\right)^2$$

The following relations between the average small variations $\Delta q_1$, $\Delta q_2$, $\Delta q_3$ are given by :

$$\Delta q_1 = \frac{q_{1,max}}{N(q_{1,max})} \qquad \Delta q_2 = \frac{q_{2,max}}{N_{q_1}(q_2)} \qquad \Delta q_3 = \frac{q_{3,max}}{N_{q_1,q_2}(q_3)},$$

where $q_{i,max}$ is the maximum value of coordinate $q_i$, $N(q_1)$ is the number of $q_1$ values selected randomly between 0 and $q_1$, $N_{q_1}(q_2)$ is the number of $q_2$ values selected randomly for a given value of $q_1$ and $N_{q1,q2}(q_3)$ is the number of $q_3$ values selected randomly for given values of $q_1$ and $q_2$.

A uniform distribution in the $q_1 q_2 q_3$ space will be obtained if:

$$h_1 \Delta q1 = h_2 \Delta q_2 = h_3 \Delta q_3$$

since these three displacements are orthogonal to each other. Then, we can deduce the relations between the three numbers of random points $N(q_1)$, $N_{q1}(q_2)$, $N_{q1,q2}(q_3)$ as we have done for spherical coordinates.

## Conclusion

Dealing with a problem of circular or spherical symmetry, it looks *a priori* natural to use polar or spherical coordinates. Remind you for instance the treatment of angular momenta in Quantum Mechanics, the study of the motion of planets around the Sun or the temporal deformation of the Sun surface which exhibits s, p, d,… motions. Think also to the decomposition of scattering wave functions in partial waves. Moreover, the use of non Cartesian coordinates can be justified to point out some specific symmetry properties which does not appear immediately to the eyes using Cartesian coordinates.

Nevertheless it is clear that caution should be considered to guarantee a correct pictorial description. Moreover, this analysis gives a pedagogical lightening on MC methodology. Of course, this analysis is not restricted to the atomic and molecular representations. Its interest appears in any situation where Monte Carlo representations are chosen or required.

Our practical advice: If you like to overcome difficulties, enjoy playing with non Cartesian coordinates adapted to your specific problem. Now, if you want to avoid technical difficulties or bad surprises, prefer the use of Cartesian coordinates. It is less exciting, but you will get correct results in total security.

## ■ AUTHOR INFORMATION
\* corresponding author e-mail: jc.rayez@ism.u-bordeaux1.fr

The authors thank deeply M. T. Rayez for her help to improve the presentation.

## ■ REFERENCES and FOOTNOTE

(1) Cromer D.T., Stereo Plots of Hydrogen-like Electron Densities. J. Chem. Edu. **1968**, 45, 626 – 631.

(2) Look at for instance i) "Wolfram MathWorld, the web's most extensive mathematics resource" http://mathworld.wolfram.com/SpherePointPicking.htm (Sphere point picking)
ii) http://kanga.usask.ca/Geant/Exercises/sampling2.html (Sampling in non-cartesian coordinate systems)
iii) http://mragheb.com/NPRE%20498MC%20Monte%20Carlo%20Simulations%20in%20Engineering/Sampling%20Special%20Distributions.pdf (Sampling special distributions" by M. Ragheb in Monte Carlo Simulation NPRE 498MC)

(3) The number 31714 obtained numerically above can be easily deduced analytically. $N_r(\theta)$ is given by: $N_r(\theta) = N(r_{max}) \cdot \dfrac{2\pi.r}{r_{max}}$ [formula 2]. The uniform distribution of r values on the distance $r_{max}$ leads to $\Delta r = \dfrac{r_{max}}{N(r_{max})}$ [formula 1]. The values of r can be linked to $\Delta r$ by: $r = n.\Delta r$ with n varying from 1 to $N(r_{max})$. Then:

$$\sum_{r=0}^{r_{max}} N_r(\theta) = 2\pi \cdot \frac{N(r_{max}).\Delta r}{r_{max}} \cdot \sum_{n=1}^{N(r_{max})} n = 2\pi \cdot \sum_{n=1}^{N(r_{max})} n = 2\pi \cdot \frac{N(r_{max}).(N(r_{max})+1)}{2} = \pi.100 \times 101$$

Taking $\pi = 3.14$ without any other digit (which is consistent with the number 628 in the text), we get the value 31714 exactly.

(4) Eyring H., Walter J., Kimball G.E. *Quantum Chemistry*, ed. by John Wiley and Sons, New-York, 14th printing 1967, p. 363 -368.

**Figures captions**

**Figure 1a** – Uniform distribution of 31714 dots in a circular disk of area 100 Å$^2$. The number of random θ points are linked to the number of random r points according to relation [2] in the text. The value of 100 Å$^2$ for the disk area has been chosen since above 5 Å the density of the $2p_x$ function is very small.

**Figure 1b** – Non-uniform distribution of 10000 dots obtained in the same circular disk as in Figure 1a, but with a random selection of the same number (100 each) of r and θ points.

**Figure 1c** – Uniform distribution of 40380 dots in a square of side length equal to the diameter of the circular disk of area 100 Å$^2$.

**Figure 2a** – $2p_x$ density function distribution obtained from the 31714 random (r, θ) dots displayed in Figure 1a. 2982 dots only remained after application of the Monte Carlo condition.

**Figure 2b** – $2p_x$ density function distribution obtained from the 10000 random (r, θ) dots displayed in Figure 1b. Only 5407 dots remained at the end.

**Figure 2c** – $2p_x$ density function distribution obtained from the 40380 random (x, y) dots displayed in Figure 1c. Only 3034 dots are present.

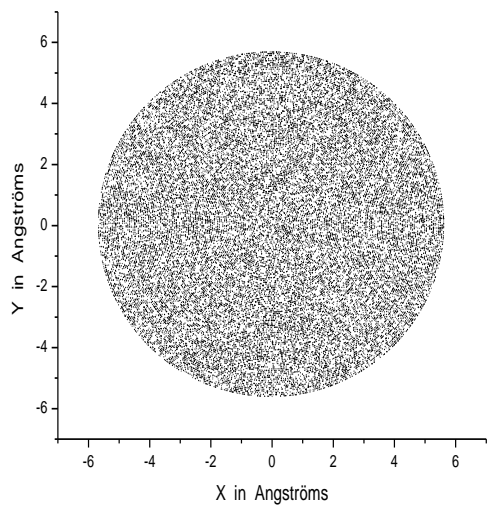

**Figure 1a**

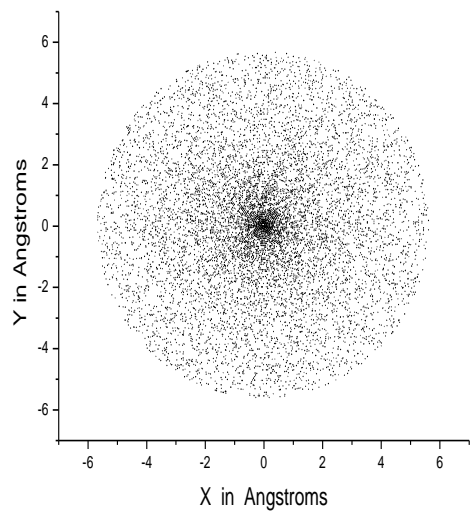

**Figure 1b**

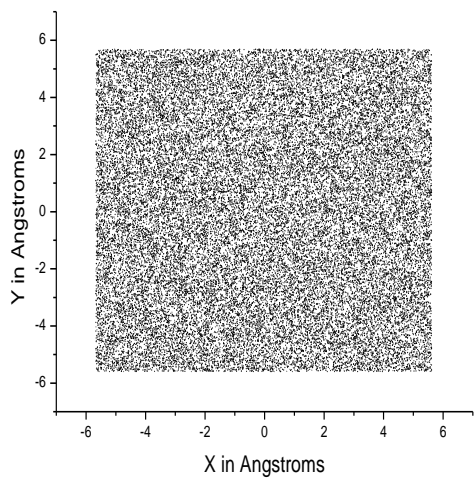

**Figure 1c**

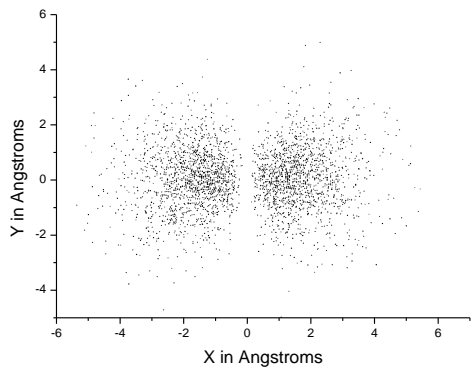

**Figure 2a**

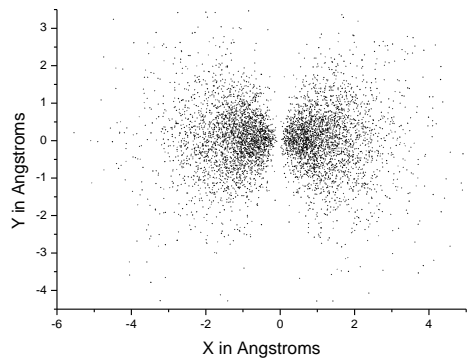

**Figure 2b**

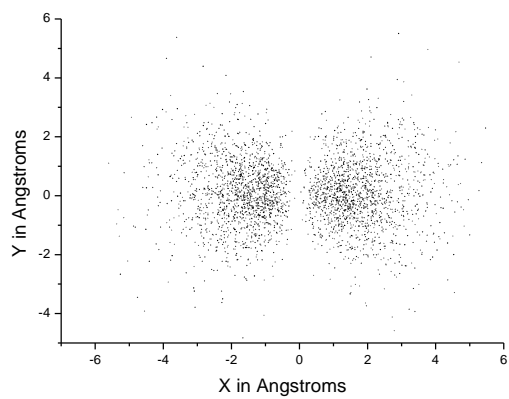

**Figure 2c**